\documentclass[review]{elsarticle}

\usepackage{graphicx}
\usepackage{dcolumn}
\usepackage{bm}
\usepackage{epstopdf}
\usepackage{color}
\usepackage{url}
\usepackage{hyperref}

\begin{document}

\begin{frontmatter}

\title{A simple and efficient kinetic model for wealth distribution with saving propensity effect: based on lattice gas automaton}

\author[Address1]{Lijie Cui}
\author[Address2,Address3]{Chuandong Lin\corref{mycorrespondingauthor}}
\cortext[mycorrespondingauthor]{Corresponding author}
\ead{linchd3@mail.sysu.edu.cn}

\address[Address1]{School of Labor Economics, Capital University of Economics and Business, Beijing 100070, China}
\address[Address2]{Sino-French Institute of Nuclear Engineering and Technology, Sun Yat-Sen University, Zhuhai 519082, China}
\address[Address3]{Key Laboratory for Thermal Science and Power Engineering of Ministry of Education, Department of Energy and Power Engineering, Tsinghua University, Beijing 100084, China}

\begin{abstract}
The dynamics of wealth distribution plays a critical role in the economic market, hence an understanding of its nonequilibrium statistical mechanics is of great importance to human society. For this aim, a simple and efficient one-dimensional (1D) lattice gas automaton (LGA) is presented for wealth distribution of agents with or without saving propensity. The LGA comprises two stages, i.e., random propagation and economic transaction. During the former phase, an agent either remains motionless or travels to one of its neighboring empty sites with a certain probability. In the subsequent procedure, an economic transaction takes place between a pair of neighboring agents randomly. It requires at least $4$ neighbors to present correct simulation results. The LGA reduces to the simplest model with only random economic transaction if all agents are neighbors and no empty sites exist. The 1D-LGA has a higher computational efficiency than the 2D-LGA and the famous Chakraborti-Chakrabarti economic model. Finally, the LGA is validated with two benchmarks, i.e., the wealth distributions of individual agents and dual-earner families. With the increasing saving fraction, both the Gini coefficient and Kolkata index (for individual agents or two-earner families) reduce, while the deviation degree (defined to measure the difference between the probability distributions with and without saving propensities) increases. It is demonstrated that the wealth distribution is changed significantly by the saving propensity which alleviates wealth inequality. 
\end{abstract}

\begin{keyword}
Lattice gas automaton \sep Agent-based model \sep Wealth distribution \sep Wealth inequality \sep Saving propensity
\MSC[2010] 	62P20\sep 65C20 \sep 68U20
\end{keyword}

\end{frontmatter}


\section{Introduction}

In econophysics, various economic and financial issues can be analyzed and solved with probabilistic methods of statistical physics \cite{Stanley1996PA,Mantegna1999,Yakovenko2009RMP}. As an open problem in economics and econophysics, the fundamental dynamics of wealth distribution has been widely studied due to its key role in human as well as nonhuman society \cite{Kohler2017Nature,Acemoglu2002QJE,Chase2020PA}. For the description of wealth distribution, a famous analytical tool is the Boltzmann-Gibbs exponential function \cite{Dragulescu2001} (see Eq. (\ref{Probability1})), another well-known empirical approach is the Pareto power-law function $P\left( m \right)\propto {{m}^{-\alpha }}$ in terms of the Pareto index $\alpha$ and individual wealth $m$ \cite{Cardoso2020PA}. Historical data indicate that the Boltzmann-Gibbs function is usually reasonable for the low and middle ranges of wealth distribution \cite{Dragulescu2001,Tao2019JEIC}, while the Pareto function provides a good fit to the high range \cite{Nirei2007RIW,Newby2011EM}. The Gaussian-like distribution which is observed for the lower wealth of the population (around $90 \%$) is due to the additive process of lower wealth accumulation \cite{Cardoso2020PA,Vallejos2018JEIC}. While the power-law tail for the top ($10 \%$ or so) is because of the multiplicative way of the wealth in a higher group \cite{Cardoso2020PA,Vallejos2018JEIC}. Recently, to further explain the datasets for $67$ countries, Tao et al. presented a theoretical model within the standard framework of modern economics and demonstrated that free competition and Rawls' fairness are the underlying mechanisms producing the exponential pattern \cite{Tao2019JEIC}.

Besides the aforementioned analytical and empirical studies, numerical research provides convenient insight into the wealth distribution with the emergence of various versatile computational methods \cite{Dragulescu2000EPJB,Chakraborti2000EPJB,Bourguignon2006JEI,Pareschi2014PTRSA,Juan2018DGA,Alves2019CNSNS}. In 2000, Dr$\rm{\breve{a}}$gulescu and Yakovenko presented both analytical arguments and computational simulations for the exponential distribution that emerges in computer simulations of economic models, and discussed the role of debt and models with broken time-reversal symmetry for which the Boltzmann-Gibbs law does not hold \cite{Dragulescu2000EPJB}. In the same year, the Chakraborti-Chakrabarti (CC) economic model was proposed for a closed economic system with a fixed number of agents and total money, and the saving propensity influence upon the statistical mechanics of wealth distribution was studied \cite{Chakraborti2000EPJB}. In 2006, Bourguignon and Spadaro reviewed microsimulation techniques and their theoretical background as a tool for the investigation of public policies \cite{Bourguignon2006JEI}. In 2014, Pareschi and Toscani extended a nonlinear kinetic equation of Boltzmann type that describes the effect of knowledge on the wealth of agents who interact through binary trades \cite{Pareschi2014PTRSA}. In 2018, an agent-based model was considered to investigate the wealth distribution where the interchange was determined with a symmetric zero-sum game \cite{Juan2018DGA}. In 2019, Alves and Monteiro modified a spatial evolutionary version of the ultimatum game as a toy model suitable for wealth distribution \cite{Alves2019CNSNS}. 

As an effective stochastic methodology, the lattice gas automaton (LGA) is a simple kinetic model that is applicable to the hydrodynamics \cite{FHP1986PRL}, chemistry \cite{Chen2016Entropy}, electromagnetics \cite{Simons1999}, thermoacoustics \cite{Chen2004JAP}, and economics \cite{Cerda2013MCM,Cui2020Entropy}, etc. The LGA was pioneered by the Hardy-Pomeau-de Pazzis model \cite{HPP1973} and the later Frisch-Hasslacher-Pomeau model \cite{FHP1986PRL}. In 2013, Cerd$\rm{\acute{a}}$ et al. presented a two-dimensional (2D) LGA for income distribution in a market with charity regulations \cite{Cerda2013MCM}. Very recently, a modified LGA economic model was developed for the income distribution under the conditions of the Matthew effect, income tax and charity \cite{Cui2020Entropy}. In fact, the LGA is based on the mechanism that there is a one-to-infinite mapping between a macroscopic performance and various microscopic details, thus the realistic phenomenon can be manifested by a collective group of artificial particles evolving on lattices in an appropriate way \cite{LGABook2005,SucciBook}. This idea also enlightened the development of other versatile methodologies, such as the lattice Boltzmann method (LBM) \cite{SucciBook,McNamara1988,Qian1992EL,Lai2014PA,Li2019ENTROPY} and discrete Boltzmann method (DBM) \cite{Lin2017PRE,Lin2018CNF,Lin2019CTP,Lin2019PRE,Gan2015SM,Lai2016PRE,Gan2018PRE,Gan2019FOP,Ye2020Entropy}. Actually, these mesoscopic kinetic models (including the LGA, LBM, DBM) have attracted great attention due to their simple schemes, flexible applications, easy programing, and high parallel computing efficiency, etc. 

Motivated by previous investigations \cite{Chakraborti2000EPJB,Cerda2013MCM,Cui2020Entropy}, an effective 1D-LGA is proposed for wealth distribution in an economic society where people have saving propensities or not. Compared with the 2D-LGA \cite{Cerda2013MCM,Cui2020Entropy}, the current model is simpler and faster, and the factor of individual saving propensity is taken into account as well. Moreover, the LGA has a higher computational efficiency than the famous CC-model \cite{Chakraborti2000EPJB}. The rest of the paper is organized as follows. In Sec. \ref{SecII}, the LGA economic model is introduced in detail. In Sec. \ref{SecIII}, the model is validated and then used to study the wealth distribution with saving propensity effect. Finally, conclusions are drawn in Sec. \ref{SecIV}. 

\section{Lattice gas automaton}\label{SecII}

In practice, the wealth distribution of agents in a closed free market takes the form of an exponential Boltzmann-Gibbs function, which is analogous to the energy distribution in statistical physics \cite{Yakovenko2009RMP}. Here, the LGA is constructed to describe an artificial society where a monetary exchange may occur if two agents encounter after random movements. Similarly to statistical physics, the agents, wealth and human society are equivalent to the ideal gas molecules, internal energy and particle system, respectively. 

\begin{figure}[tbp]
	\begin{center}
		\includegraphics[bbllx=0pt,bblly=0pt,bburx=271pt,bbury=288pt,width=0.4\textwidth]{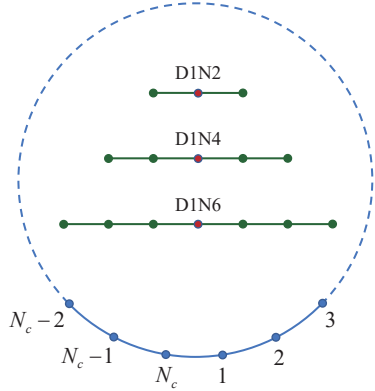}
	\end{center}
	\caption{Computational domain with $N_{c}$ sites and discrete models (D1N2, D1N4 and D1N6).}
	\label{Fig01}
\end{figure}

Let us consider a simple economic system where the number of agents $N_{a}$ is fixed and the total amount of money $M$ is conserved. An agent $A_{i}$ that represents an individual or a corporation owns money $m_{i}$, with the subscript $i = 1$, $2$, $\cdots$, $N_{a}$. Initially, the total money $M$ is divided amongst $N_a$ agents, hence each agent possesses the same amount of money $m_{0} = M / N_{a}$. All agents are randomly located in a circle with sites $N_{c}$ (under the condition $N_{c} \geq N_{a}$), see Fig. \ref{Fig01}. The spatial and temporal steps are $\Delta x = 1$ and $\Delta t = 1$, respectively. 

In the evolution of the LGA, there are two key stages, i.e., the random propagation and economic transaction. 

\textbf{Stage 1: Propagation}

An agent can move to its neighboring empty sites (with probability $P_{m}$) or keep resting (with probability $1-{P_{m}}$) in the stage of random propagation. For simplicity, the left and right neighboring sites are symmetrical, namely, the number of neighbors is even. For example, there are $1$ left and $1$ right neighboring sites in the model $1$-dimensional-$2$-neighbor (D1N2), while there are $2$ neighboring positions on each side in D1N4, see Fig. \ref{Fig01}. Hence, there are $2$, $4$, and $6$ neighboring sites for D1N2, D1N4, and D1N6, respectively. Numerical tests show that it needs at least $4$ neighbors for the LGA to obtain right simulation results, see Fig. \ref{Fig02}. That is to say, the D1N2 model presents incorrect results while the D1N4 and D1N6 are satisfactory. 

\textbf{Stage 2: Transaction}

In the phase of economic transaction, two neighboring agents $A_{i}$ and $A_{j}$ trade with probability $P_{t}$. (For example, an agent may deal with one of $4$ neighbors in D1N4.) Each agent's money is always non-negative, namely, no debt is permitted. Conservation of the total money is obeyed in each exchange, as earlier. An arbitrary pair of agents $A_i$ and $A_j$ get engaged in an exchange with trading volume $\Delta m$, i.e.,
\begin{equation}
\left\{
\begin{array}{l}
{{m}'_{i}}={{m}_{i}}-\Delta m \tt{,}  \\
{{m}'_{j}}={{m}_{j}}+\Delta m \tt{,}
\end{array}
\right.
\label{TradeI}
\end{equation}
where ${m}_{i}$ and ${m}'_{i}$ are the money amounts of $A_i$ before and after the transaction, and similar to ${m}_{j}$ and ${m}'_{j}$ for $A_j$. In this work, two types of trade models are adopted for agents with or without saving propensity \cite{Chakraborti2000EPJB,Cerda2013MCM,Cui2020Entropy} (see \ref{APPENDIXA} for details).

Remark: For the case $N_{c} > N_{a}$, the sequence could be ``Propagation + Transaction" or ``Transaction + Propagation" in the main loop of the program; For the special case $N_{c} = N_{a}$, there is no empty site, so all agents remain motionless and no propagation takes place. In the latter case, the LGA becomes a reduced model with only economic transaction. The LGA is extremely robust and independent of a specific initial condition, see \ref{APPENDIXB}.

\section{Numerical simulations}\label{SecIII}

In theory \cite{Dragulescu2001}, for arbitrary and random trades with local money conservation in a market, the wealth distribution approaches the equilibrium Boltzmann-Gibb distribution of statistical mechanics. It is proved that the stationary wealth distribution functions of individual agents and two-earner families in an ideal free market take the Dr$\rm{\breve{a}}$gulescu-Yakovenko (DY) forms \cite{Dragulescu2001}, 
\begin{equation}
{{P}_{1}}\left( m \right)=\frac{1}{{{m}_{0}}}\exp \left( -\frac{m}{{{m}_{0}}} \right)
\label{Probability1}
\tt{,}
\end{equation}
and 
\begin{equation}
P_{2} \left( m \right) = \frac{m}{m_{0}^{2}} \exp \left(- \frac{m}{m_{0}} \right)
\label{Probability2}
\tt{,}
\end{equation}
respectively. Despite its simplicity, the theoretical model misses a very natural ingredient for realistic transactions: Almost no economic agent exchanges with the entire wealth without saving some parts; The saving propensity is a natural tendency for any normal economic agent. This defect also exists in the trade model-I without saving propensity, which is equivalent to the trade model-II with saving propensity in the case $\lambda = 0$.

In the following subsections, we first consider the above simple case, i.e., transaction without saving propensity, and compare simulation results to exact solutions in Eqs. (\ref{Probability1}) and (\ref{Probability2}). Next, the transaction with saving propensity is taken into account. 

\subsection{Transaction without saving propensity}

\begin{figure}[tbp]
	\begin{center}
		\includegraphics[bbllx=0pt,bblly=0pt,bburx=575pt,bbury=201pt,width=0.99\textwidth]{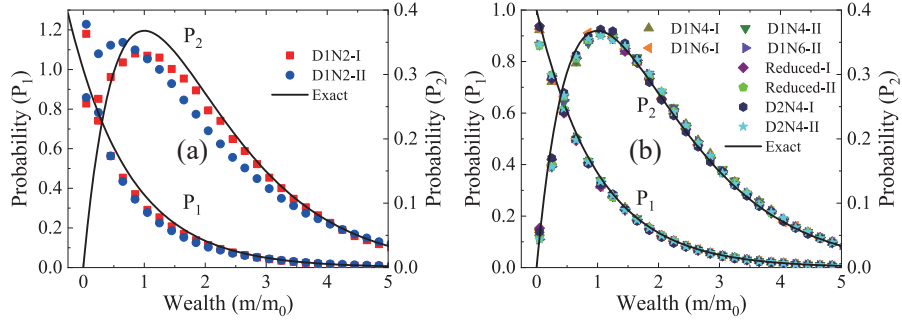}
	\end{center}
	\caption{Wealth distributions of individual agents (on the left axis) and dual-earner families (on the right axis), respectively. The symbols denote the simulation results of D1N2-I (squares), D1N2-II (circles), D1N4-I (upper triangles), D1N4-II (lower triangles), D1N6-I (left triangles), D1N6-II (right triangles), Reduced-I (diamonds), Reduced-II (pentagons), D2N4-I (hexagons), and D2N4-II (stars). The lines represent the corresponding exact solutions.}
	\label{Fig02}
\end{figure}

The site number is chosen as $N_{c} = 1500$ for D1N4 and D1N6, and $N_{c} = 600$ for the reduced model. The other parameters are $P_{m} = 0.8$, $P_{t} = 0.7$, $N_{A} = 600$, $m_{0} = 1$. Figure \ref{Fig02} illustrates the wealth distributions of individual agents and dual-earner families, respectively. In the legend, for convenience, ``I" refers to the first trade model without saving propensity, and ``II" refers to the second trade model with saving propensity in the case $\lambda = 0$. In Fig. \ref{Fig02} (a), the squares and circles indicate the simulation results of D1N2 using the first trade model (D1N2-I) and the second one (D1N4-II), respectively. In Fig. \ref{Fig02} (b), the upper, lower, left and right triangles stand for D1N4-I, D1N4-II, D1N6-I, and D1N6-II, respectively. The diamonds and pentagons are for the reduced model (without propagation) using the first and second trade models, respectively. Besides, the two-dimensional models, D2N4-I (diamonds) and D2N4-II (stars), are used as well \cite{Cerda2013MCM,Cui2020Entropy}. Meanwhile, the solid lines are for the corresponding exact solutions in Eqs. (\ref{Probability1}) and (\ref{Probability2}). 

It is apparent in Fig. \ref{Fig02} that the market is non-interacting and the resulting individual wealth distribution takes the equilibrium Boltzmann-Gibb form. Most agents own little wealth, the maximum probable money is zero, and the population becomes lower for a larger fortune. Meanwhile, the density of dual-earner families firstly increases then reduces with increasing wealth, and the maximum is located at $m = m_{0}$. Figure \ref{Fig02} (a) depicts that the simulation results of D1N2-I and D1N2-II have a relatively large departure from the exact solutions (\ref{Probability1}) and (\ref{Probability2}). While all numerical results in Fig. \ref{Fig02} (b) agree well with the exact solutions (\ref{Probability1}) and (\ref{Probability2}). It is confirmed that, except D1N2, both 1D- and 2D-LGA could present the correct simulation results of wealth distribution in an ideal free market. The trade model-I is consistent with the trade model-II for $\lambda = 0$. In addition, the average money exactly remains constant $m_{0}$ during all simulations, which demonstrates that the money conservation is guaranteed in the LGA. 

\begin{table}[h]
	\centering
	\begin{tabular}{lcccc}
		\hline\hline
		Model & Temporal step & Interval & Relaxation time & Computing time \\
		\hline
		D1N4-I & $5 \times 10^{5}$ & $1000$ & $20 \times 1000$ & $208$ s \\
		\hline
		D1N4-II & $500$ & $1$ & $12$ & $9$ s \\
		\hline
		D1N6-I & $5 \times 10^{5}$ & $1000$ & $13 \times 1000$ & $332$ s \\
		\hline
		D1N6-II & $500$ & $1$ & $6$ & $9$ s \\
		\hline
		Reduced-I & $5000$ & $10$ & $10 \times 10$ & $218$ s \\
		\hline
		Reduced-II & $500$ & $1$ & $1$ & $36$ s \\
		\hline
		D2N4-I & $5 \times 10^{5}$ & $1000$ & $30 \times 1000$ & $438$ s \\
		\hline
		D2N4-II & $500$ & $1$& $8$ & $10$ s \\
		\hline\hline
	\end{tabular}
	\caption{Computational time (steps) taken by various models}
	\label{TableI}
\end{table}

It should be mentioned that the simulated smooth stationary distributions in Fig. \ref{Fig02} are determined by an average over a sequence of dynamic probabilities at set intervals. Table \ref{TableI} shows the temporal step, interval, relaxation time and computing time taken by those models in Fig. \ref{Fig02} (b). From table \ref{TableI} the following points can be obtained. 

(i) The number of probability distributions is $n_{d} = t_{s} / t_{b}$ in terms of the temporal step $t_{s}$ and interval $t_{b}$. Clearly, there are $n_{d} = 500$ sets of probability distributions in each simulation. Note that the distributions under consideration are in near equilibrium states after an early relaxation process, during which the economic system starts to approach the (near) equilibrium state from an initial configuration \cite{Chakraborti2000EPJB,Cui2020Entropy}. 

(ii) Numerical tests show that the LGA has a very high computational efficiency. For example, to conduct the above simulations, it only takes $208$, $332$, and $218$ seconds (s) for models D1N4-I, D1N6-I, and Reduced-I, respectively. (The computing time has a narrow variation for different runs of the same program due to its random nature.) Here the computational facility is a personal computer with Intel(R) Core(TM) i7-8750H CPU @ 2.20 GHz and RAM 16.0 GB. 

(iii) Model-I takes more relaxation time and computing time than model-II. For example, the relaxation time is $2 \times 10^{4}$ and $12$ temporal steps for D1N4-I and D1N4-II, respectively. The running time needs $208$ and $9$ s for D1N4-I and D1N4-II, respectively. The reason is that the trade volume in the first trade model (without saving propensity) is smaller than the mean exchange volume in the second trade model (with saving propensity). Consequently, the latter model is more efficient than the former one. 

(iv) The 1D-models have less relaxation time and computing time than the 2D-models. For instance, the relaxation time is $2 \times 10^{4}$ and $3 \times 10^{4}$ temporal steps for D1N4-I and D2N4-I, respectively. The running time needs $208$ and $438$ s for D1N4-I and D2N4-I, respectively. In other words, the 1D-model has a higher calculation efficiency than the 2D-model. Additionally, it is obvious that the former is simpler than the latter in the program as well. 

(v) It takes longer computing time per temporal step and shorter relaxation time for a model with more neighbors. The computing time per temporal step is $t_{p} = t_{c} / t_{s}$ with the computing time $t_{c}$ and temporal step $t_{s}$. For example, the results are $t_{p} = 4.2 \times {10}^{-4}$ s and $t_{p} = 6.6 \times {10}^{-4}$ s for D1N4-I and D1N6-I, respectively. The relaxation time is $t_{r} = 2 \times 10^{4}$ and $1.3 \times 10^{4}$ temporal steps for D1N4-I and D2N6-I, respectively. 

(vi) It requires less temporal steps and intervals for a model with more neighbors. The computing time is almost the same for two models with different neighbors to achieve the (near) equilibrium state. For example, the computing time within the relaxation process is $t_{r} \times t_{p} = 8.4$ s and $8.6$ s for D1N4-I and D1N6-I, respectively. This is because there are more economic transactions and longer running time for a model with more neighbors during one main loop of the program. And the artificial economic system requires approximately the same transaction times to reach an equilibrium state. 

\subsection{Transaction with saving propensity}

\begin{figure}[tbp]
	\begin{center}
		\includegraphics[bbllx=0pt,bblly=0pt,bburx=592pt,bbury=477pt,width=0.9\textwidth]{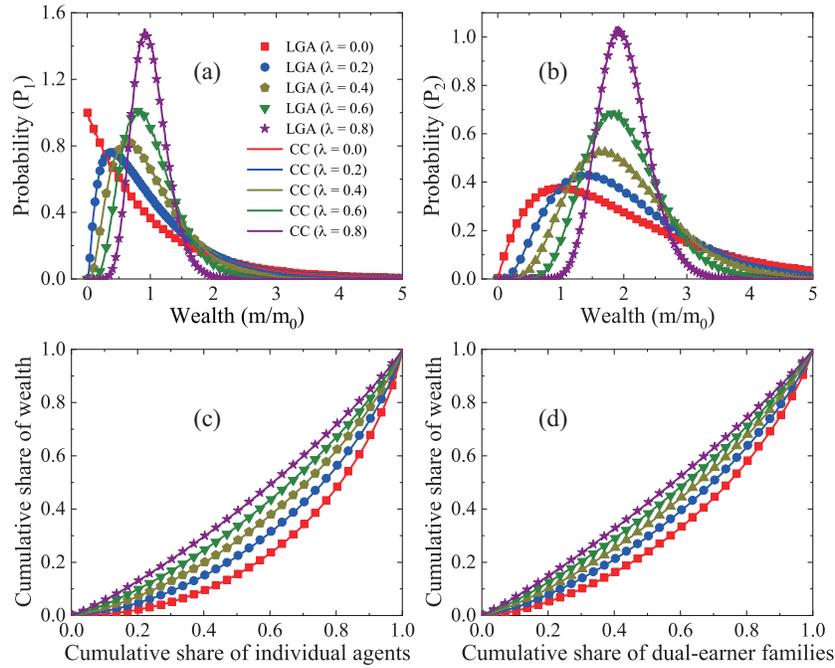}
	\end{center}
	\caption{Wealth distributions of individual agents (a) and dual-earner families (b), and cumulative wealth shares of individual agents (c) and dual-earner families (d) with various saving fractions. The symbols denote the LGA results with $\lambda = 0.0$ (squares), $0.2$ (circles), $0.4$ (pentagons), $0.6$ (triangles), and $0.8$ (stars), respectively. The lines represent the corresponding CC-model results \cite{Chakraborti2000EPJB}.}
	\label{Fig03}
\end{figure}

Now, let us consider the wealth distribution under the condition of individual saving propensity. Figure \ref{Fig03} displays the wealth probability distributions and cumulative wealth shares with various saving fractions from $\lambda = 0.0$ to $0.8$. The symbols indicate the simulation results of the LGA, whose parameters are the same as those of D1N4-II in Fig. \ref{Fig02} (b). The solid lines denote the corresponding results of the CC-model \cite{Chakraborti2000EPJB}. The two sets of numerical results coincide exactly with each other. It is numerically verified that the saving propensity of agents is incorporated appropriately within the LGA. 

In addition, the saving propensity destroys the multiplicative property of the distributions in Eqs. (\ref{Probability1}) and (\ref{Probability2}). As shown in Fig. \ref{Fig03} (a), the wealth distribution changes from the Boltzmann-Gibb form to the asymmetric Gaussian-like form with a finite $\lambda$ introduced. The agents with zero wealth gradually decrease and even disappear with the increasing $\lambda$. A peak of $P_{1}$ emerges as $\lambda$ is large enough. Figure \ref{Fig03} (b) shows that there is a peak of $P_{2}$ for any $\lambda$. For either $P_{1}$ or $P_{2}$, the peak becomes thinner and higher, and moves rightward for a larger saving fraction. It can be found in Figs. \ref{Fig03} (c) and (d) that the cumulative shares of wealth rise from $0$ to $1$ as the cumulative shares of either individual agents and dual-earner families increase from $0$ to $1$. With increasing saving fractions, the curve of the cumulative shares approaches the linear equality line and its curvature becomes lower.

Furthermore, it takes about $21$ and $88$ s for the LGA and CC-model to conduct the above simulations, respectively. Namely, the computational cost of the CC-model is about four times that of the LGA. Hence, the LGA has a higher computational efficiency than the CC-model \cite{Chakraborti2000EPJB}, although both are quite simple and efficient. It is reasonable, because the random trade is between two neighboring agents subsequently from $i = 1$ to $N_{a}$ in the LGA, while the pair undertaking the transaction are chosen in an arbitrary and random way from all $N_{a}$ agents in the CC-model \cite{Chakraborti2000EPJB}. Therefore, the LGA requires less (mean) times for that all agents have traded. 

To measure the wealth inequality under saving propensity, we introduce the Gini coefficient expressed by
\begin{equation}
G = \frac{1}{2N_{a}^{2}{{w}_{0}}}\sum\nolimits_{i=1}^{{{N}_{w}}}{\sum\nolimits_{j=1}^{{{N}_{w}}}{\left| {{w}_{i}}-{{w}_{j}} \right|}}
\tt{,}
\end{equation}
which theoretically ranges from $0$ (complete equality) to $1$ (complete inequality). Specifically, the Gini coefficient $G_{1}$ depends on the parameters $w_{0} = m_{0}$, $N_{w} = N_{a}$, $w_{i} = m_{i}$ for individual agents; And $G_{2}$ is a function of $w_{0} = 2 m_{0}$, $N_{w} = N_{a} / 2$, $w_{i} = m_{i}+m_{i+N_{a}/2}$ for two-earner families. 

Apart from the Gini coefficient, another important parameter is the Kolkata ($k$) index that gives an intuitive measure of wealth inequality \cite{Ghosh2014PA,Chatterjee2017PA}. It is defined as follows: $1-k$ fraction of population possess top $k$ fraction of wealth in the society, namely, the cumulative wealth of $1-k$ fraction of people exceed those earned by the rest $k$ fraction of the people \cite{Ghosh2014PA,Chatterjee2017PA}. The $k$ index is from $0.5$ (complete equality) to $1$ (complete inequality). Furthermore, Kolkata index can be rescaled to unit interval via the transformation $K=2k-1$ which gives the vertical distance between the perfect equality line and the Lorentz curve at the point $k$. 

Besides, to describe the departure of probability distribution with saving propensity from that without saving propensity (i.e., the DY form (\ref{Probability1}) or (\ref{Probability2})), we define the deviation degree as
\begin{equation}
\Delta 
= \frac{1}{2} \int_{0}^{\infty }{\left| f - f^{eq} \right|} dm
\tt{,}
\end{equation}
which is between $0$ (complete overlap) and $1$ (no overlap), see \ref{APPENDIXC} for more details. In particular, the symbols $f$ and $f^{eq}$ denote the wealth distributions for an arbitrary value of $\lambda$ and $\lambda = 0$, respectively. Namely, $f = P_{1} (\lambda)$ and $f^{eq} = P_{1} (\lambda = 0)$ for individual agents; $f = P_{2} (\lambda)$ and $f^{eq} = P_{2} (\lambda = 0)$ for two-earner families. 

\begin{figure}[tbp]
	\begin{center}
		\includegraphics[bbllx=0pt,bblly=0pt,bburx=295pt,bbury=632pt,width=0.45\textwidth]{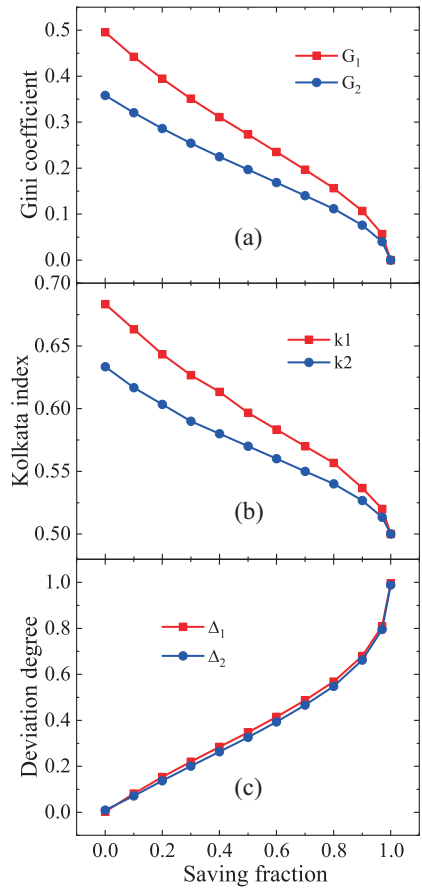}
	\end{center}
	\caption{The Gini coefficients (a), Kolkata indices (b), and deviation degrees (c) versus saving fractions. The lines with squares and circles are for individual agents and two-earner families, respectively.}
	\label{Fig04}
\end{figure}

Figure \ref{Fig04} (a) plots the Gini coefficients versus saving fractions. The lines with squares and circles represent the Gini coefficients $G_{1}$ for individual agents and $G_{2}$ for two-earner families, respectively. It is apparent that the Gini coefficient $G_{1}$ for individual agents is not lower than $G_{2}$ for dual-earner families. The Gini coefficients $G_{1}$, $G_{2}$, and their differences are smaller for a larger saving fraction. Compared to the theoretical solutions $G_{1} = 1/2$ and $G_{2} = 3/8$ for $\lambda = 0$ \cite{Dragulescu2001}, the corresponding calculation results $G_{1} = 0.499$ and $G_{2} = 0.371$ are satisfactory. And the LGA results $G_{1} = G_{2} = 0$ coincide well with the exact solutions at the point $\lambda = 0$ \cite{Dragulescu2001}. Moreover, it is interesting to obtain the relationship $3 G_{1} \approx 4 G_{2}$ for all saving fractions. 

Figure \ref{Fig04} (b) illustrates the Kolkata indices versus saving fractions. The lines with squares and circles denote the Kolkata indices $k_{1}$ and $k_{2}$ for individual agents and dual-earner families, respectively. Obviously, the Kolkata index $k_{1}$ for individual agents is greater than or equal to $k_{2}$ for dual-earner families. The Kolkata indices $k_{1}$, $k_{2}$, and their differences decrease with the increasing saving fraction. In comparison with the analytic solutions $k_{1} = 0.682156$ and $k_{2} = 0.634555$ for $\lambda = 0$ \cite{Dragulescu2001,Ghosh2014PA,Chatterjee2017PA}, the corresponding simulation results $k_{1} = 0.683$ and $k_{2} = 0.633$ are satisfying. Meanwhile, the simulation results $k_{1}=k_{2}=0.5$ are in line with the exact solutions at the point $\lambda = 1$ \cite{Dragulescu2001,Ghosh2014PA,Chatterjee2017PA}. Additionally, comparison between Figs. \ref{Fig04} (a) and (b) indicates a linear relationship between the Gini coefficients and Kolkata indices, i.e., $k_{1} = 0.5 + \gamma_{1} G_{1}$, $k_{2} = 0.5 + \gamma_{2} G_{2}$, and $\gamma_{1} \approx \gamma_{2} \approx 0.36$, which are close to the results in Ref. \cite{Chatterjee2017PA}. Similar to the Gini coefficients, the indices $K_{1}=2k_{1}-1$ for individual agents and $K_{2}=2k_{2}-1$ for dual-earner families satisfy the relation $3 K_{1} \approx 4 K_{2}$ within $0 \le \lambda \le 1$. 

At last, Fig. \ref{Fig04} (c) illustrates the deviation degrees versus saving fractions. The line with squares indicates the deviation degree $\Delta _{1}$ that describes the departure of the wealth distribution of individual agents from the DY expression (\ref{Probability1}); The line with circles is for the deviation degree $\Delta _{2}$ which measures the difference between the wealth distribution of two-earner families and the DY formula (\ref{Probability2}). The calculation results $\Delta _{1} = 0.004$ and $\Delta _{2} = 0.009$ for $\lambda = 0$ are satisfactory by comparison with the corresponding theoretical solutions $\Delta _{1} = \Delta _{2} = 0$. Meanwhile, the numerical results $\Delta _{1} = 0.9951$ and $\Delta _{1} = 0.9886$ agree well with the exact ones $\Delta_{1} = \Delta _{2} = 1$ at the point $\lambda = 1$. Remarkably, with the increasing saving fraction $\lambda$, the deviation degrees $\Delta _{1}$ and $\Delta _{2}$ increase. That is to say, the wealth inequality is alleviated by the saving propensity, and the wealth distribution is affected by the human factor. Additionally, it is interesting to find the relation $\Delta _{1} \approx \Delta _{2}$ (with only slight differences) between them within the whole range of $\lambda$. 

\section{Conclusion}\label{SecIV}

We proposed a quite simple, robust and effective kinetic method, 1D-LGA, for the wealth distribution in a closed economic market where the amount of money and the number of agents are fixed. Analogously to statistical physics, the agents, wealth and human society are equivalent to the ideal gas molecules, internal energy and particle system, respectively. The LGA includes two key stages, i.e., random propagation and economic transaction. During the propagation stage, an agent either remains motionless or travels to one of its neighboring empty sites with a certain probability. In the subsequent procedure, an economic transaction takes place randomly when two agents are located in the neighboring sites. Two types of transaction models are introduced. One is model-I for agents without saving propensity \cite{Cerda2013MCM,Cui2020Entropy}, the other is model-II with saving propensity \cite{Chakraborti2000EPJB}. The former is equivalent to the latter if the saving fraction is zero. 

Numerical tests indicate that to obtain right simulation results requires at least four neighbors. The LGA reduces to the simplest coarse-grained model with only random economic transaction if all agents are neighbors and no empty sites exist. For a model with more neighbors, it takes longer computing time per temporal step, shorter relaxation time, less temporal steps and intervals. However, the total computing time is almost the same for two models with different neighbors to achieve the (near) equilibrium state. Because there are more economic transactions and longer running time for a model with more neighbors during one main loop of the program. And the artificial economic system requires approximately the same transaction times to obtain an equilibrium state. Furthermore, model-I takes more relaxation time and computing time than model-II, because the trade volume in the former is smaller than the mean exchange volume in the latter. Consequently, the latter model is more efficient than the former one. The 1D-LGA is more efficient and simpler than the 2D-LGA \cite{Cerda2013MCM,Cui2020Entropy}, and also takes less computing time than the CC-model \cite{Chakraborti2000EPJB}, although all these models have a quite high computational efficiency. 

Next, the LGA is validated with two benchmarks, i.e., the wealth distributions of individual agents and two-earner families. The LGA is extremely robust and independent of a specific initial condition. It presents the numerical results of wealth distributions with various saving propensity factors exactly the same as the CC-model \cite{Chakraborti2000EPJB}. To be specific, the wealth distribution changes from the Boltzmann-Gibb form to the asymmetric Gaussian-like form with a finite $\lambda$ introduced. The agents with zero wealth gradually decrease and even disappear with the increasing $\lambda$. A peak of individual wealth distribution emerges as $\lambda$ is large enough, while there is a peak of wealth distribution of two-earner families for any $\lambda$. For either of them, the peak becomes thinner and higher, and moves rightward for a larger saving fraction. It is noteworthy that the LGA has the potential to describe the main feature of the wealth distribution in human society.

Finally, the Gini coefficient and Kolkata index are used to measure wealth inequality under saving propensity. Meanwhile, the deviation degree is defined to describe the departure of probability distribution with saving propensity from that without saving propensity. With the increasing $\lambda$, the Gini coefficients $G_{1}$ for individual agents and $G_{2}$ for two-earner families decrease, the Kolkata indices $k_{1}$ for individual agents and $k_{2}$ for two-earner families reduce, while the deviation degrees $\Delta _{1}$ for individual agents and $\Delta _{2}$ for two-earner families increase. The transformation $K_{1}=2k_{1}-1$ and $K_{2}=2k_{2}-1$ are introduced.  It is interesting to find the relations $3 G_{1} \approx 4 G_{2}$, $3 K_{1} \approx 4 K_{2}$, and $\Delta _{1} \approx \Delta _{2}$. The Gini coefficients and Kolkata indices satisfy the linear relations $k_{1} = 0.5 + \gamma_{1} G_{1}$ and $k_{2} = 0.5 + \gamma_{2} G_{2}$ with $\gamma_{1} \approx \gamma_{2} \approx 0.36$, which are similar to the results in Ref. \cite{Chatterjee2017PA}. It is demonstrated that the wealth inequality is alleviated by the saving propensity, and the wealth distribution is influenced by the human factor.

\section*{Acknowledgments}
This work is supported by the National Natural Science Foundation of China (NSFC) under Grant No. 51806116. 

\appendix

\section{}\label{APPENDIXA}

Here two types of transaction models are introduced for agents with or without saving propensity, respectively. 

\textbf{(i) Trade model-I without saving propensity}

The trading volume between two agents is fixed as $\Delta m = m_{0} / {{N}_{m}}$. Most of our simulations are for $m_{0} = 1$ and ${{N}_{m}} = 100$. There are three cases of exchange under consideration \cite{Cerda2013MCM,Cui2020Entropy}. 

Case A: $m_{i} = m_{j} = 0$. No exchange takes place between two agents without personal possessions. 

Case B: $m_{i} = 0$ and $m_{j} \neq 0$. An agent without any wealth can only stay unchanged or win money during an economic exchange. 

Case C: $m_{i} \neq 0$ and $m_{j} \neq 0$. An agent either earns or loses money with probability $P_{t}/2$. 

\textbf{(ii) Trade model-II with saving propensity}

Assume that each economic agent saves a fraction $\lambda$ of its wealth $m_{i}$ before trading, where $\lambda$ is a fixed value between zero and unity. The parameter $\lambda$, also called the ``marginal propensity to save", remains fairly constant, independent of economic agents \cite{Chakraborti2000EPJB}. After the transaction, the wealth of agents $A_i$ and $A_j$ becomes 
\begin{equation}
\left\{
\begin{array}{l}
{{m}'_{i}}={\lambda}{{m}_{i}}+{\Delta m _{i}} \tt{,}  \\
{{m}'_{j}}={\lambda}{{m}_{j}}+{\Delta m _{j}} \tt{,}
\end{array}
\right.
\label{TradeII}
\end{equation}
in terms of $\Delta {{m}_{i}}=\epsilon \left( 1-\lambda  \right)\left( {{m}_{i}}+{{m}_{j}} \right)$ and $\Delta {{m}_{j}}=\left( 1-\epsilon  \right)\left( 1-\lambda  \right)\left( {{m}_{i}}+{{m}_{j}}\right)$, where $\epsilon$ represents a random number between zero and unity \cite{Chakraborti2000EPJB}. Via straight-forward substitution, it can be derived that Eq. (\ref{TradeII}) is equivalent to Eq. (\ref{TradeI}) for $\Delta m = \left( 1-\lambda  \right)\left[ {{m}_{i}}-\varepsilon \left( {{m}_{i}}+{{m}_{j}} \right) \right]$. The random exchange amount is less than the total money because of the saving by each agent.

\section{}\label{APPENDIXB}

\begin{figure}[tbp]
	\begin{center}
		\includegraphics[bbllx=0pt,bblly=0pt,bburx=336pt,bbury=243pt,width=0.5\textwidth]{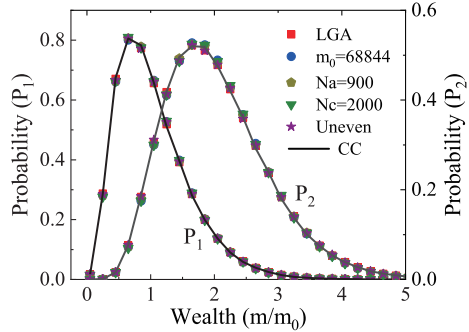}
	\end{center}
	\caption{Wealth distributions of individual agents (on the left axis) and dual-earner families (on the right axis) for $\lambda = 0.4$. The symbols denote the simulation results of the LGA for a different $m_0$, $N_a$, $N_{c}$, and an uneven configuration. The lines represent the corresponding CC-model results \cite{Chakraborti2000EPJB}.}
	\label{Fig05}
\end{figure}

It is worth mentioning that the LGA is extremely robust and independent of a specific initial configuration. For this purpose, Fig. \ref{Fig05} delineates the wealth distributions of individual agents and dual-earner families. The squares and lines correspond to the LGA and CC-model results for $\lambda = 0.4$ in Fig. \ref{Fig03}, respectively. The circles, pentagons, and triangles stand for the LGA results for $m_{0} = 68844$, $N_{a} = 900$ and $N_{c} = 2000$, respectively. The stars are for an uneven initial configuration, $m_{i} = m_{0} + A_{0} \sin \left(2 \pi i /{N_{a}} \right)$, with $i = 1$, $2$, $\cdots$, $N_{a}$. Here the perturbation amplitude $A_{0} = m_{0} / 2$ is utilized to generate an unequal wealth distribution of agents. It is clear that all simulation results agree well with each other. Consequently, the stationary distribution is not related to the average wealth, the number of agents, the number of sites, or the initial wealth distribution. 

\section{}\label{APPENDIXC}

\begin{figure}[tbp]
	\begin{center}
		\includegraphics[bbllx=0pt,bblly=0pt,bburx=583pt,bbury=235pt,width=0.99\textwidth]{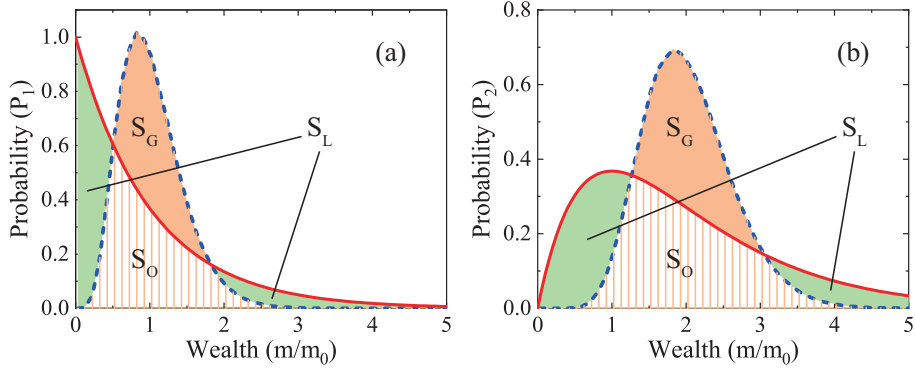}
	\end{center}
	\caption{Sketch of the wealth distributions of individual agents (a) and dual-earner families (b). The solid lines stand for $\lambda = 0$, and the dotted lines for $\lambda \ne 0$.}
	\label{Fig06}
\end{figure}

Here we introduce a useful parameter to describe the departure of the wealth distribution with saving propensity from the one without saving propensity. To give an intuitive description, Fig. \ref{Fig06} delineates the sketch of the wealth distributions of individual agents (a) and dual-earner families (b). The solid lines stand for the case without saving propensity, and the dotted lines for the other case. The area below each line equals one because of the following formulas,
\begin{equation}
S_{f} = \int_{0}^{\infty }{f } dm = 1
\tt{,}
\end{equation}
\begin{equation}
S_{f^{eq}} = \int_{0}^{\infty }{f^{eq} } dm = 1
\tt{.}
\end{equation}
The overlap between the two regions $S_{f}$ and $S_{f^{eq}}$ is $S_{O}$, and the areas are $S_{L}$ and $S_{G}$ for $f < f^{eq}$ and $f > f^{eq}$, respectively. To be specific,
\begin{equation}
{{S}_{L}}={{\left. \int_{0}^{\infty }{\left( {{f}^{eq}}-f \right)}dm \right|}_{f < {{f}^{eq}}}}
\tt{,}
\end{equation}
\begin{equation}
{{S}_{G}}={{\left. \int_{0}^{\infty }{\left( f - {{f}^{eq}} \right)}dm \right|}_{f > {{f}^{eq}}}}
\tt{.}
\end{equation}
Then, the relation $S_{L} + S_{O} = S_{G} + S_{O} = 1$ leads to 
\begin{equation}
S_{L} = S_{G} = \frac{1}{2} \int_{0}^{\infty }{\left| f - f^{eq} \right|} dm
\tt{.}
\end{equation}
The deviation degree is defined as the ratio of nonoverlapping area to total area, i.e.,
\begin{equation}
\Delta =\frac{{{S}_{L}}+{{S}_{G}}}{{{S}_{f}}+{{S}_{{{f}^{eq}}}}}=\frac{\int_{0}^{\infty }{\left| f-{{f}^{eq}} \right|}dm}{\int_{0}^{\infty }{\left| f+{{f}^{eq}} \right|}dm}
= \frac{1}{2} \int_{0}^{\infty }{\left| f - f^{eq} \right|} dm
\tt{,}
\end{equation}
whose the range is $0 \le \Delta \le 1 $. To be specific, the case $\Delta = 0$ refers to $f = f^{eq}$, namely, the wealth distributions $f$ and $f^{eq}$ coincide with each other; The case $\Delta = 1$ correspond to the circumstance that there is no overlap between the two areas. With $\Delta$ increasing from zero to one, the distribution $f$ departs far and far from $f^{eq}$. 

\section*{References}

\bibliography{References}

\end{document}